\documentclass[epsf]{article}
\usepackage{graphics}

\makeatletter

\providecommand{\LyX}{L\kern-.1667em\lower.25em\hbox{Y}\kern-.125emX\@}
\let\SF@@footnote\footnote
\def\footnote{\ifx\protect\@typeset@protect
    \expandafter\SF@@footnote
  \else
    \expandafter\SF@gobble@opt
  \fi
}
\expandafter\def\csname SF@gobble@opt \endcsname{\@ifnextchar[%]
  \SF@gobble@twobracket
  \@gobble
}
\edef\SF@gobble@opt{\noexpand\protect
  \expandafter\noexpand\csname SF@gobble@opt \endcsname}
\def\SF@gobble@twobracket[#1]#2{}

\newcommand{\bee}{\begin{equation}}
\newcommand{\ee}{\end{equation}}
\newcommand{\beea}{\begin{eqnarray}}
\newcommand{\eea}{\end{eqnarray}}

\makeatother

\begin{document}

\thispagestyle{empty} \parskip=12pt \raggedbottom

\noindent \def\mytoday\#1  \ifcase\month \or
 January\or February\or March\or April\or May\or June\or
 July\or August\or September\or October\or November\or December\fi
\space \number\year{} \hspace*{9cm} COLO-HEP-441\\
 \hspace*{9.3cm} December 1999.

\noindent \vspace*{1cm}

{\par\centering {\large Spatial Correlation of the Topological Charge in Pure
SU(3) Gauge Theory and in QCD}\large \par}

{\par\centering \vspace{0.5cm}\par}

{\par\centering Anna Hasenfratz\\
 Physics Department, University of Colorado, \\
 Boulder, CO 80309 USA\par}

\begin{abstract}
{\par\centering We study the spatial correlator of the topological charge density
operator in pure SU(3) gauge theory and in two flavor QCD. We show that the
data for distances up to about 1 fm is consistent with a vacuum consisting of
individual instantons and closely bound pairs. The percentage of paired objects
is twice as large on the dynamical configurations than on the pure gauge ones,
implying increased molecule formations due to fermionic interactions.\par}
\end{abstract}
\eject

Our understanding of the QCD vacuum has increased considerably in the last few
years. Lattice calculations using different algorithms have consistent predictions
for the topological susceptibility in pure gauge QCD \cite{Teper_lat99}, and
the first results in dynamical systems have been published recently \cite{Teper_Hart},\cite{Pisa99}.
Results concerning the density and size distribution of instantons are less
consistent, but there is increasing evidence that the pure gauge QCD vacuum
is filled with instantons of average radius \( \sim 0.3 \) fm, with a density
of about 1 fm\( ^{-4} \) \cite{Shuryak_Schafer},\cite{SU3INST},\cite{Deforcrand_size}. 

Instanton Liquid Models (ILM) describe the phenomenological properties of instantons
in the QCD vacuum closely \cite{Shuryak_Schafer}. Even the Random ILM provides
an accurate description of the pure gauge instanton vacuum indicating that the
gauge interaction and consequently the spatial correlation between instantons
is small. The situation is quite different for systems with dynamical light
quarks. In the zero quark mass limit the topological susceptibility is zero
and there are no unpaired topological objects in the vacuum. Yet one expects
instantons to be present. Chiral symmetry is spontaneously broken, \( <\bar{\psi }\psi >\ne 0 \)
in the zero quark mass limit. According to the Casher-Banks formula, the chiral
condensate (summed over quark flavors) is \cite{Banks_Casher}

\begin{equation}
<\bar{\psi }\psi >=\pi \rho (0),
\end{equation}
implying that \( \rho (0) \), the density of the eigenmodes of the Dirac operator
at eigenvalue zero, is finite. Since instantons are the leading candidates to
create near-zero eigenmodes of the Dirac operator, the vacuum is likely to be
filled with instantons. Fermions create an attractive force between oppositely
charged instantons. It is therefore natural to expect that the instantons of
the zero quark mass vacuum form instanton-antiinstanton pairs, molecules. In
the case of finite quark mass the topological susceptibility does not vanish
but it is proportional to the quark mass \cite{Leutwyler_Smilga}
\begin{equation}
\label{eq:Leutwyler_Smilga}
\chi =<\bar{\psi }\psi >\frac{m_{q}}{n^{2}_{f}}+O(m_{\pi }^{4})=\frac{f_{\pi }^{2}m_{\pi }^{2}}{4n^{}_{f}}+O(m_{\pi }^{4}),
\end{equation}
 where \( f_{\pi }=132 \)MeV is the pion decay constant. The vacuum in this
case has unpaired objects in addition to the molecules. 

Even though the above picture describing spatial correlation between instantons
sounds very natural, no evidence from lattice calculations has supported it
so far. The culprit is most likely the lattice approach. In order to reveal
individual topological objects, vacuum fluctuations have to be removed, the
vacuum has to be smoothed. Almost all smoothing procedures distort the vacuum,
they destroy molecules especially easily. In addition the smoothed configurations
are frequently analyzed by a pattern-recognition algorithm to identify individual
instantons. Most algorithms have a built-in cut-off which limits the nearest
objects it can resolve, further limiting the possibility of finding closely
bound pairs. 

In this paper we study the topological density correlator 
\begin{equation}
\label{eq:Cz}
C(r)=\frac{1}{V}\int d^{4}x\, q(x)q(x+z),\qquad r=|z|,
\end{equation}
 where \( q(x) \) is the topological charge density measured using the improved
charge operator of Ref. \cite{SU3INST},\cite{SU2INST}. We measured \( C(r) \)
on two sets of configurations with similar lattice spacings. The first set is
a pure gauge ensemble of \( 16^{3}\times 32 \) configurations at \( \beta =6.0 \),
the is a dynamical ensemble of the same size, generated with two flavors of
staggered fermions at \( \beta =5.7 \) with \( ma=0.01 \). The lattice spacing
of the pure gauge ensemble is \( a\cong 0.095 \) fm, of the dynamical ensemble
is \( a\cong 0.11 \) fm. Both ensembles are from the NERSC QCD archive\cite{NERSC_QCD}.
For this study 170 of the pure gauge configurations \cite{Quenched_lat} and
83 of the dynamical configurations \cite{Dynamical_lat} (the entire available
data set) were used. 

The gauge configurations have to be smoothed to reveal the topological content.
Here we used APE smearing \cite{APEblock} with parameter \( c=0.45 \) as the
properties of this smoothing has been extensively tested in Refs. \cite{SU3INST},\cite{SU2INST}.
\( N \) steps of APE smearing smoothes a configuration to distance \( d_{s}\sim a\sqrt{Nc/3} \)
\cite{Fat_pert} but preserves the properties of the vacuum at longer distances.
First we will compare the topological density correlators on the pure gauge
and dynamical configurations after the same number, \( N=30 \), APE steps.
Since the lattice spacings of the two ensembles are approximately the same,
that corresponds to about the same physical smoothing and even observables that
change with smoothing can be compared. Afterwards we will vary \( N \) between
10 and 60 to justify the above assumption. 

First we consider the topological susceptibility. The susceptibility is largely
independent of the number of smoothing steps on both configuration sets. Several
previous studies found that the topological charge has very large autocorrelation
time on dynamical configurations, sometimes hundreds of molecular dynamics time
units. We did not find this problem here. The average charge on the dynamical
configurations is \( <Q>=0.16\pm0 .25 \) and the charge distribution also appears
standard. We found \( \chi ^{1/4}=193(4) \)MeV on the pure gauge ensemble and
\( \chi ^{1/4}=130(5) \)MeV on the dynamical ensemble. This decrease of the
susceptibility is expected according to Eq. \ref{eq:Leutwyler_Smilga}. Using
the published pion mass value \( m_{\pi }a=0.25 \) of the dynamical ensemble
\cite{Dynamical_lat} in Eq. \ref{eq:Leutwyler_Smilga} we obtain \( f_{\pi }=(105\pm 20) \)
MeV, fairly close to the experimental value \( f_{\pi }=132 \) MeV. (Note that
Eq. \ref{eq:Leutwyler_Smilga} differs from the equivalent equation of Ref.\cite{Teper_Hart}.\footnote{%
The author is indebted to Peter Hasenfratz for checking Eq. \ref{eq:Leutwyler_Smilga}.
})
\begin{figure}
{\par\centering \resizebox*{8.5cm}{!}{\includegraphics{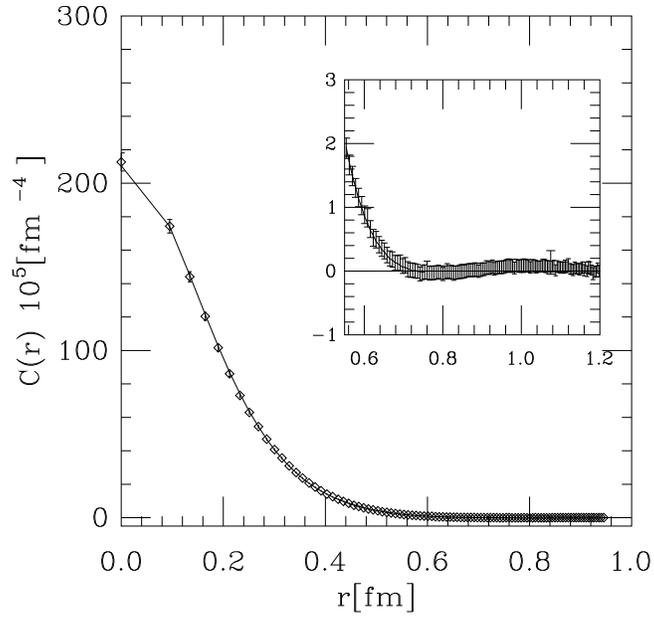}} \par}

\caption{Spatial correlation of the topological charge density on pure gauge configurations.
The insert enlarges the large \protect\( r\protect \) region. The solid line
corresponds to the four parameter fit described in the text.\label{fig:Cz_vs_z_Q}}
\end{figure}
\begin{figure}
{\par\centering \resizebox*{8.5cm}{!}{\includegraphics{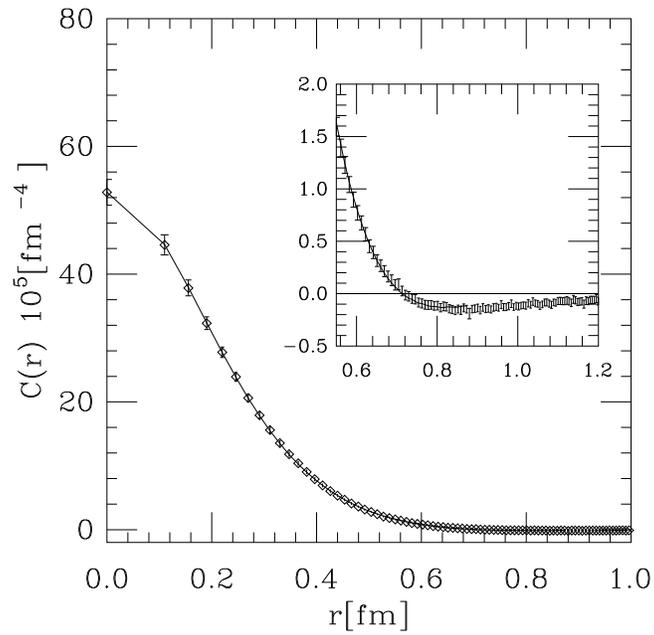}} \par}

\caption{Same as Figure \ref{fig:Cz_vs_z_Q} but for the dynamical ensemble. \label{fig:Cz_vs_z_D}}
\end{figure}

Figures \ref{fig:Cz_vs_z_Q} and \ref{fig:Cz_vs_z_D} show \( C(r) \) versus
\( r \) for both configuration sets after \( N=30 \) APE smoothing steps.
The inserts of the figures enlarge the large \( r \) tail. At this smoothing
level the vacuum fluctuations are largely removed. A few properties of the instanton
vacuum can be immediately deduced from the figures. At small distances \( C(r) \)
is dominated by the auto-correlator of individual instantons and \( C(r) \)
can be approximated by a dilute gas. In the dilute gas approximation the height
of the correlator is proportional to the number of topological objects of the
configuration,\( N_{0} \), and the width is related to the average radius of
the instantons, \( \bar{\rho } \). At distances \( r\simeq 2\bar{\rho } \),
the correlator probes the close neighbors of the instantons. If there is no
spatial correlation between topological objects, \( C(r) \) should be about
zero for \( r\geq 2\bar{\rho } \)\footnote{%
In the continuum reflexion positivity requires \( C(r)<0 \) for all \( r\ne 0 \).
On the lattice that is not the case for small \( r \). Since smoothing removes
most of the vacuum fluctuations one expects \( C(r) \) to be dominated by non-perturbative
vacuum structures.
}. If pairing of oppositely charged objects occur, \( C(r) \) is expected to
be negative, reaching its minimum around \( r\sim 2\bar{\rho } \). Comparing
now the pure gauge and dynamical correlators of figures \ref{fig:Cz_vs_z_Q}
and \ref{fig:Cz_vs_z_D} we observe that the widths of the correlators are almost
identical but the height at \( r=0 \) is very different. This implies that
\( \bar{\rho } \) is approximately the same for the two ensembles but the dynamical
configurations have about three times fewer topological objects. The tails of
the two distributions reveal further differences. While \( C(r) \) on the pure
gauge ensemble is consistent with zero for \( r\geq 0.7fm \), \( C(r) \) on
the dynamical ensemble shows an unmistakable negative dip signaling opposite-charge
correlation. 

\begin{table}

\caption{The results of the four parameter fit for the pure gauge and the dynamical
configurations. }
{\centering \begin{tabular}{|c|c|c|}
\hline 
&
Pure gauge&
Dynamical\\
\hline 
\hline 
\( N_{0}/V \) {[}fm\( ^{-4} \){]}&
1.08\( \pm  \)0.05&
0.35\( \pm  \) 0.02\\
\hline 
\( \bar{\rho } \){[}fm{]}&
0.25\( \pm  \)0.01&
0.30\( \pm  \) 0.01\\
\hline 
\( d \){[}fm{]}&
0.53\( \pm  \) 0.03&
0.61\( \pm  \) 0.02\\
\hline 
\( 2n_{p}/N_{0} \) &
0.16\( \pm  \) 0.02&
0.29\( \pm  \) 0.03\\
\hline 
\end{tabular}\par}
\label{table:fitparameters}
\end{table}

To quantify these differences, we model the vacuum with a simple picture. Assume
that there are \( N_{0} \) topological objects in the vacuum with uniform radius
\( \bar{\rho }. \) If the topological density of a single instanton of radius
\( \bar{\rho } \) centered at the origin is \( q_{0}(x) \) and the instantons
are non-overlapping, the topological density of the configuration in this model
is
\begin{equation}
\tilde{q}(x)=\sum ^{N_{0}}_{i=1}s_{i}q_{0}(x-x_{i}),
\end{equation}
where \( x_{i} \) is the center of the the ith topological object and \( s_{i}=\pm 1 \)
depending on whether it is an instanton or antiinstanton. The topological correlator
is
\begin{equation}
\tilde{C}(z)=\sum _{i,j}s_{i}s_{j}C_{\rho }(z-x_{i}+x_{j})
\end{equation}
 where \( C_{\rho }(z) \) is the topological correlator of a single instanton.
If the instantons are randomly distributed, averaging over configurations will
give 
\begin{equation}
<\tilde{C}(z)>=N_{0}C_{\rho }(z).
\end{equation}
If only \( N_{0}-n_{p} \) objects are randomly distributed and the other \( n_{p} \)
objects form molecules (i.e. \( 2n_{p} \) oppositely charged instantons are
paired), the average of the topological correlator is 
\begin{equation}
<\tilde{C}(z)>=N_{0}C_{\rho }(z)-n_{p}C_{d}(z),
\end{equation}
 where \( C_{d}(z) \) is the correlator of an instanton-instanton pair separated
by distance \( d \), averaged over the direction of \( d \). \( C_{\rho } \)
and \( C_{d} \) can be calculated using the analytical form of the single instanton
charge distribution. If the instanton size is \( \bar{\rho }\sim 0.3 \) fm
and the pairs are closely bound, i.e. \( d\sim 0.6 \) fm, this model can be
valid in the \( 0\leq r\leq 0.6-0.9 \) fm range. The measured \( C(r) \) then
can be fitted with the four parameters \( N_{0}, \) \( n_{p} \), \( \bar{\rho } \)
and \( d \). In the fit we keep data with \( 0\leq r/a\leq 5-8 \) as for \( r/a\geq L/2=8 \)
finite size effects become important. Varying the upper range of the fit between
\( r/a=5 \) and 8 hardly effects the fit parameters, it changes the \( \chi ^{2} \)
of the fit only. The solid lines in figures \ref{fig:Cz_vs_z_Q} and \ref{fig:Cz_vs_z_D}
correspond to the best fit. The fit parameters listed in table \ref{table:fitparameters}
confirm our earlier observations. The average instanton size on both ensembles
is about \( 0.3 \) fm. The instanton density on the pure gauge ensemble is
about \( 1 \) fm\( ^{-4} \), in agreement with expectations, while the density
on the dynamical configurations is about a third of that. On both ensembles
we observe pairs at a distance of about \( 2\bar{\rho } \). On the pure gauge
configurations 15 percent of the instantons are in pairs, while on the dynamical
configurations close to 30 percent are paired up. It is not surprising to find
pairs even on the pure gauge configurations, as there is an attractive gauge
interaction between oppositely charged instantons, but pairing is clearly enhanced
on the dynamical ensemble.

Until now we have compared the dynamical and pure gauge ensembles at the same
level of APE smoothing. In the remainder of this paper we study the dependence
on the level of smoothing. The smoothing has two different effects on a configuration:
it removes vacuum fluctuations and it also distorts and annihilates instantons.
These two effects cannot be separated. The first effect is the important one
for small amounts of smoothing, but as the configuration is smoothed many times,
the second one becomes more relevant. We have measured the correlator \( C(r) \)
at \( N=10,20,30,40,50 \) and \( 60 \) levels of APE blocking on both ensembles
and also at \( N=15 \) on the dynamical configurations only. The same level
of APE smoothing can have very different effect at different lattice spacing.
The smearing distance \( d_{s}=a\sqrt{Nc/3} \) characterizes the removal of
vacuum fluctuations but it is not clear what is the variable (if one exists
at all) that describes how instantons are annihilated or otherwise distorted
by APE smoothing at different lattice spacings. In the following we report the
results as the function of the smearing distance. The qualitative form of the
correlator is the same at all levels of smearing and we fit the data with four
parameters as described above. The fit is excellent in every case except at
the \( N=60 \) level smoothing where the smearing distance is comparable to
the lattice size and finite size effects become important. 
\begin{figure}
{\par\centering \resizebox*{8.5cm}{!}{\includegraphics{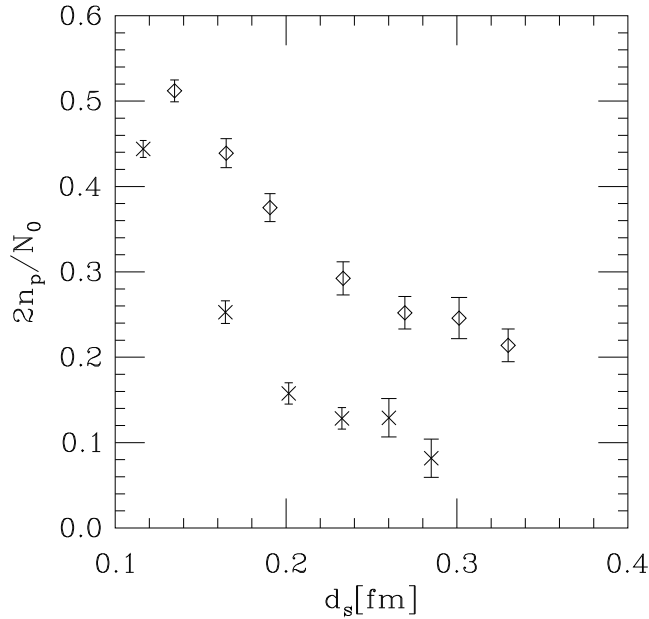}} \par}

\caption{\protect\( 2n_{p}/N_{0}\protect \) as the function of smearing distance for
the dynamical (diamonds) and pure gauge (crosses) ensembles.\label{fig:2np_over_N0}}
\end{figure}

Figure \ref{fig:2np_over_N0} shows the ratio of paired objects and \( N_{0} \)
as the function of \( d_{s} \). The ratio decreases rapidly up to \( d_{s}\sim 0.2 \)
fm but changes slower after that, indicating that most of the vacuum fluctuations
have been removed and further change is due to the annihilation of topological
objects. Comparing data at the same smoothing distance even increases the previously
observed difference between the pure gauge and dynamical ensembles; the dynamical
configurations show more than twice the pairing rate of the pure gauge ones. 

\begin{figure}
{\par\centering \resizebox*{8cm}{!}{\includegraphics{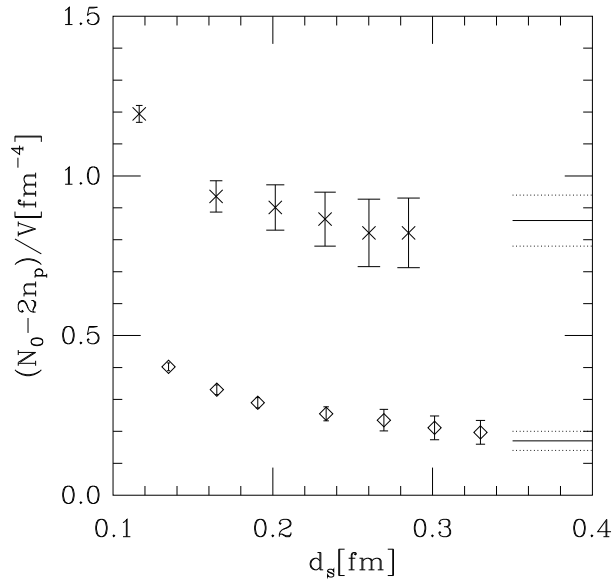}} \par}

\caption{\protect\( (N_{0}-2n_{p})/V\protect \) as the function of the smearing distance
for the dynamical (diamonds) and pure gauge (crosses) ensembles. The horizontal
lines indicate the value of the corresponding topological susceptibilities.\label{fig:N0m2np_vs_ape}}
\end{figure}

It is interesting to ask, which instantons contribute to the topological susceptibility,
all of them or only the unpaired ones? Figure \ref{fig:N0m2np_vs_ape} shows
the density of the unpaired objects, \( (N_{0}-2n_{p})/V \) as a function of
the smearing distance \( d_{s} \). The horizontal lines indicate the value
of the corresponding topological susceptibilities in fm\( ^{-4} \). The agreement
implies that the unpaired instantons follow a Poisson distribution and only
they contribute to the topological susceptibility.

We have studied only one set of dynamical ensemble with two quark flavors. It
would be important to extend this work to include different, preferably smaller
quark masses and four quark flavors. In both cases one expects the fermionic
effects to be stronger. We mention here that the NERSC archive contains two
smaller (less then 50 configurations) dynamical configuration sets, both at
heavier quark masses. We have analyzed the available configurations, but within
errors they show no significant difference compared to the dynamical configurations
discussed here. 

In summary, we have demonstrated important differences between pure gauge and
dynamical QCD configurations. On the dynamical configurations the topological
susceptibility is considerably smaller than on the pure gauge configurations.
This is due both to the decreased density of instantons and to the stronger
pairing of oppositely charged objects. In the case we studied here we found
that the density of instantons decreased by about a factor of three while pairing
occurred twice as frequently on the dynamical configurations. 

The author is indebted to T. DeGrand for carefully reading the manuscript. This
calculation would not have been possible without the configurations available
at NERSC. We would like to thank the Colorado High Energy experimental group
for allowing us to use their work stations. This work was supported by the U.S.
Department of Energy.

\newcommand{\PL}[3]{{Phys. Lett.} {\textbf{#1}} {(19#2)} #3} \newcommand{\PR}[3]{{Phys. Rev.} {\textbf{#1}} {(19#2)}  #3} \newcommand{\NP}[3]{{Nucl. Phys.} {\textbf{#1}} {(19#2)} #3} \newcommand{\NPP}[3]{{Nucl. Phys. Proc. Suppl.} {\textbf{#1}} {(19#2)} #3} \newcommand{\PRL}[3]{{Phys. Rev. Lett.} {\textbf{#1}} {(19#2)} #3} \newcommand{\PREPC}[3]{{Phys. Rep.} {\textbf{#1}} {(19#2)}  #3} \newcommand{\ZPHYS}[3]{{Z. Phys.} {\textbf{#1}} {(19#2)} #3} \newcommand{\ANN}[3]{{Ann. Phys. (N.Y.)} {\textbf{#1}} {(19#2)} #3} \newcommand{\HELV}[3]{{Helv. Phys. Acta} {\textbf{#1}} {(19#2)} #3} \newcommand{\NC}[3]{{Nuovo Cim.} {\textbf{#1}} {(19#2)} #3} \newcommand{\CMP}[3]{{Comm. Math. Phys.} {\textbf{#1}} {(19#2)} #3} \newcommand{\REVMP}[3]{{Rev. Mod. Phys.} {\textbf{#1}} {(19#2)} #3} \newcommand{\ADD}[3]{{\hspace{.1truecm}}{\textbf{#1}} {(19#2)} #3} \newcommand{\PA}[3]{ {Physica} {\textbf{#1}} {(19#2)} #3} \newcommand{\JE}[3]{ {JETP} {\textbf{#1}} {(19#2)} #3} \newcommand{\FS}[3]{ {Nucl. Phys.} {\textbf{#1}}{[FS#2]} {(19#2)} #3}

\end{document}